\documentclass{article}
\usepackage{spconf}
\usepackage{amsmath}
\usepackage{graphicx}

\usepackage{epsf}
\usepackage{times}
\usepackage{float}
\usepackage{amssymb}
\usepackage{wasysym}
\usepackage{color}
\usepackage{stfloats}
\usepackage{subeqnarray}
\usepackage{cite}
\usepackage{multirow}

\title{Attention-Driven Multichannel Speech Enhancement in Moving Sound Source Scenarios}
\name{Yuzhu Wang, Archontis Politis, Tuomas Virtanen
}
\address{ \large
\hskip -7pt \begin{tabular}{c}
Audio Research Group, Tampere University, Tampere, Finland
\end{tabular}
}

\begin{document}
\ninept
\maketitle
\begin{sloppy}
\begin{abstract}
\label{sec:abstract}
Current multichannel speech enhancement algorithms typically assume a stationary sound source, a common mismatch with reality that limits their performance in real-world scenarios. This paper focuses on attention-driven spatial filtering techniques designed for dynamic settings. Specifically, we study the application of linear and nonlinear attention-based methods for estimating time-varying spatial covariance matrices used to design the filters. We also investigate the direct estimation of spatial filters by attention-based methods without explicitly estimating spatial statistics. The clean speech clips from \textit{WSJ0} are employed for simulating speech signals of moving speakers in a reverberant environment. The experimental dataset is built by mixing the simulated speech signals with multichannel real noise from \textit{CHiME-3}. Evaluation results show that the attention-driven approaches are robust and consistently outperform conventional spatial filtering approaches in both static and dynamic sound environments.
\end{abstract}
%
\section{Introduction}
\label{sec:introduction}
Microphone arrays have gradually become an indispensable sensing front-end for future intelligent speech communication and human-machine interaction as they capture acoustic signals and preserve spatial information of the sound field \cite{benesty2008microphone,vincent2018audio}.
Despite the rapid progression in multichannel speech enhancement algorithms based on microphone arrays \cite{wang2020binaural,jin2023differential}, recovering clean speech signals in real-world noisy environments remains a significant challenge.
Recently, combining conventional signal processing with deep neural networks (DNNs) has opened new avenues to address long-standing challenges such as sound source localization, source separation, noise reduction, and de-reverberation \cite{adavanne2018sound,purwins2019deep,SIBOpan,markovic2022implicit}.

The widely accepted assumption in speech processing tasks is that target and interfering sources remain stationary during an utterance, which often deviates from real-world scenarios. Several works have explored the impact of the movement of sound sources or microphone arrays \cite{nikunen2017separation,pertila2021mobile,fujimura23_interspeech}. Speech enhancement employing spatial filtering is particularly sensitive to the spatial location of the desired source, as the motion complicates the estimation of time-varying statistics of the signal and interference.
To address the issues, existing spatial filtering solutions can be broadly categorized into three approaches: \textit{conventional}, \textit{DNN-integrated}, and \textit{fully learnable}.
Conventional multichannel spatial filtering methods compute the spatial covariance matrices (SCMs) of target and interference signals by averaging the instantaneous SCMs (ISCMs) at individual time-frequency bins \cite{kubo2019mask}.
Then, the obtained SCMs are applied to compute the multichannel spatial filters. The conventional approaches cannot precisely compute highly time-varying SCMs from sounds such as moving speakers, as the weighting is pre-determined and independent of the signal statistics.
The DNN-integrated spatial filtering is commonly a multi-stage system \cite{erdogan2016improved}, in which DNN technology is incorporated into the conventional spatial filtering framework to enhance key modules such as feature extraction \cite{heymann2016neural}, statistical estimation \cite{tan2022neural}, and modeling \cite{zhao2016dnn}.
These techniques display remarkable performance for non-moving situations, but these methods have not been evaluated on moving sources.
Recent studies incorporate the attention mechanism \cite{vaswani2017attention} into the mask-based beamforming framework to improve performance in moving source situations \cite{ochiai2023mask}. A neural network implemented with self-attention layers is used to learn the attention weights, which subsequently replace the existing SCM averaging strategies.
Fully learnable spatial filters can be constructed entirely from DNNs, eliminating the need for covariance estimation and explicit filter computation \cite{luo2019fasnet,tolooshams2020channel,koyama2020exploring}. Research indicates that such DNN-centric systems can learn to leverage spatial features implicitly and achieve competitive filtering effects \cite{casebeer2021nice}.
The performance of attention-based methods within this framework in dynamic scenarios, to the best of the authors’ knowledge, has not been explored.

Motivated by the effectiveness of the attention mechanism in the temporal sequence processing \cite{vaswani2017attention}, this paper aims to address the limitations of the existing approaches by focusing on the application of the attention mechanism in spatial filtering.
Our research revolves around the minimum variance distortionless response (MVDR) filtering structure.
Specifically, we first explore three ways of utilizing attention mechanisms to estimate the SCMs.
1) We use \cite{ochiai2023mask} as the starting point and employ a linear attention-based (LA) module to learn attention weights. Then such weights are used to determine the linear combination of ISCMs for estimating the SCMs.
2) As an extension, we utilize a nonlinear attention-based (NLA) module to estimate the SCM freely.
3) We propose a novel method for estimating scaled inverses-of SCMs to be used in an MVDR to avoid numerical instability arising from matrix inversion operations. 
We also investigate a fully learnable attention-based approach that learns the free spatial filter from the multichannel mixture, instead of estimating the SCMs.
All methodologies are implemented in a framework that operates causally and is trained in an end-to-end manner. 

We evaluate the methods in scenarios involving both static and moving sound sources. Within the DNN-integrated spatial filtering framework, the LA method outperforms others in terms of auditory evaluation metrics such as perceptual evaluation of speech quality (PESQ) \cite{rix2001perceptual} and short-time objective intelligibility (STOI) \cite{taal2011algorithm}, while the NLA method achieves the highest signal-to-distortion ratio (SDR) \cite{vincent2006performance}. 
The fully learnable spatial filtering is also proven effective in dynamic source scenarios and outperforms conventional spatial filtering.
Notably, the performance of the attention-driven approaches remains robust across dynamic and static environments. 

\section{Signal Model}
\label{sec:signal-model}
Consider a microphone array composed of $M$ elements and a speech source in a space that includes reverberation and background noise. By employing the short-time Fourier transform (STFT), the observed signal $Y_{m}(f,t)$ at microphone $m$ is represented as a superposition of the speech signal $X_{m}(f,t)$ and noise $N_{m}(f,t)$.
The signals captured by all microphones can be represented as a vector $\mathbf{y}(f, t) = [Y_1(f,t), Y_2(f,t), \ldots, Y_M(f,t)]^{T}$,
where superscript $^{T}$ denotes the transpose. 
Likewise, the speech and noise signals can be expressed as $\mathbf{x}(f, t)$ and $\mathbf{n}(f, t)$, with $\mathbf{y}(f,t) = \mathbf{x}(f,t) + \mathbf{n}(f,t)$.
The SCM of the observed signals is defined as $\mathbf{\Phi}_{\mathbf{yy}}(f,t) = \mathbb{E}[\mathbf{y}(f, t) \mathbf{y}^H(f, t)]$,
where $\mathbb{E}[\cdot]$ denotes the expected value and superscript $^H$ is the conjugate transpose. The SCMs for speech and noise signals are represented as $\mathbf{\Phi}_{\mathbf{xx}}(f,t)$ and $\mathbf{\Phi}_{\mathbf{nn}}(f,t)$ similarly. 
Speech enhancement is done by a complex-valued linear filter $\mathbf{h}(f,t)$ operating as
\begin{align}
\label{eq:multi-channel-filtering}
Z(f,t) &= \mathbf{h}^{H}(f,t) \mathbf{y}(f,t),
\end{align}
where $Z(f,t)$ is the filtered estimation.
Without the loss of generality, we choose the speech signal on the reference microphone as the desired signal, $X_{\text{ref}}(f,t) = \mathbf{u}_{\text{ref}}^{H} \mathbf{x}(f,t)$. The index of the reference microphone is given by a one-hot vector $\mathbf{u}_{\text{ref}}$ of length $M$. 
Minimization of the mean-squared error of the estimate of Eq.~\eqref{eq:multi-channel-filtering} under the speech distortionless constraint leads to the MVDR filter \cite{souden2009optimal},
\begin{equation}
\label{eq:MVDR-filter}
\mathbf{h}_{\text{MVDR}}(f,t) = \frac{\mathbf{\Phi}_{nn}^{-1}(f,t)\mathbf{\Phi}_{xx}(f,t)}
{\text{tr} [ \mathbf{\Phi}_{nn}^{-1}(f,t)\mathbf{\Phi}_{xx}(f,t) ]} \mathbf{u}_{\text{ref}},
\end{equation}
where $\text{tr}[\cdot]$ represents the trace of the matrix. The MVDR filter balances noise suppression versus speech distortion, which both are important in the perceptual quality of speech enhancement.
\begin{figure*}[htb]
  \centering
  \includegraphics[width=160mm]{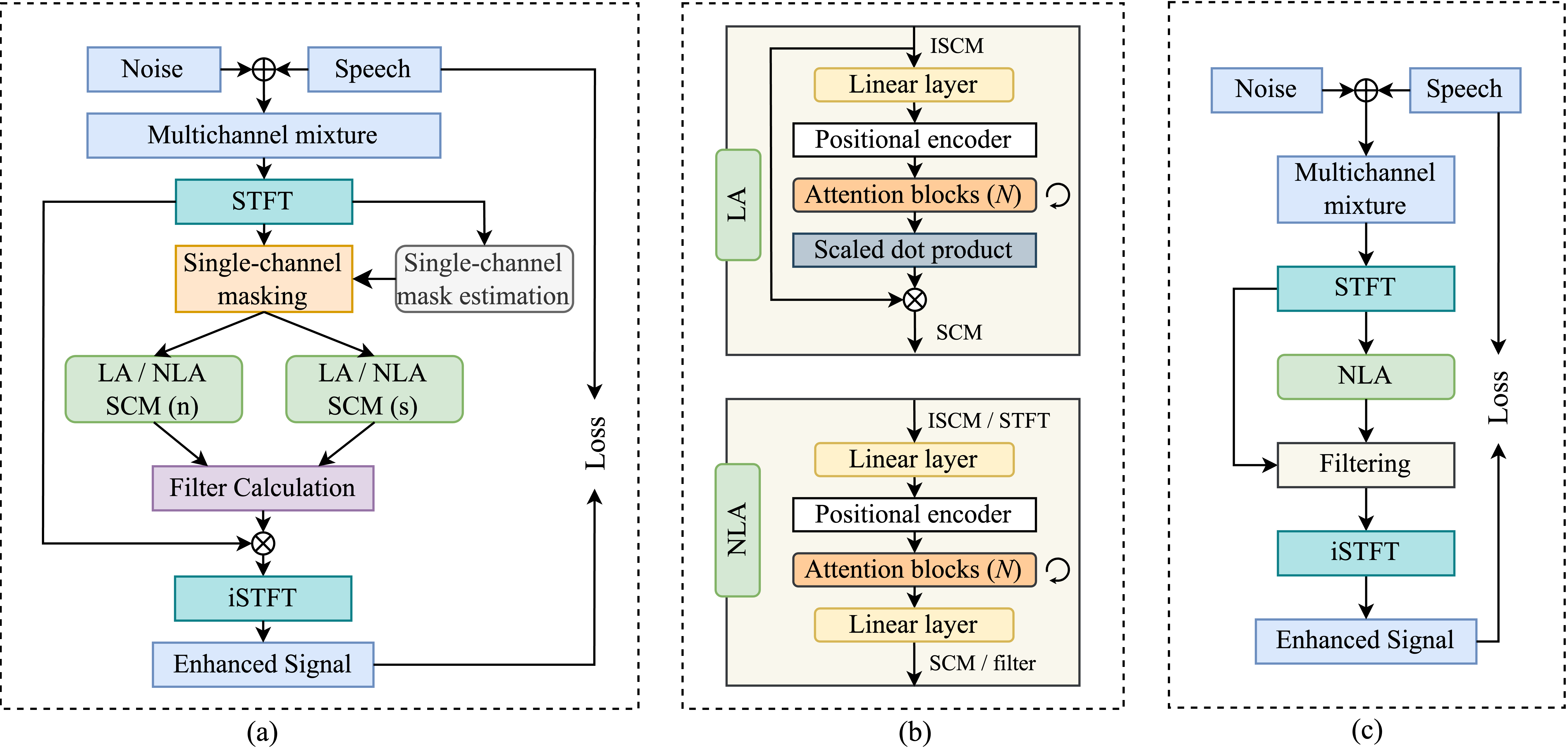}\\ \vspace{-6pt}
  \caption{(a) End-to-end framework for DNN-integrated spatial filtering. (b) LA and NLA modules. (c) End-to-end framework of fully learnable spatial filtering.
  Sharp-cornered rectangles represent numerical values or computations, rounded rectangles represent layers or networks with learnable parameters.
  }
 \vspace{-2ex}
 \label{fig:diagram}
\end{figure*}
\vspace{1ex}
\section{Attention-Driven spatial filtering}
\label{sec:AD-MCLF}
The attention-driven spatial filtering solutions applicable to moving sound sources can be applied in two frameworks: \textit{DNN-integrated spatial filtering} and \textit{fully learnable spatial filtering}. 
\subsection{DNN-integrated spatial filtering}
\label{ssec:DNN-I-MCLF}
\vspace{1ex}
\subsubsection{Framework}
\label{sssec:framework}
The DNN-Integrated spatial filtering system used in this paper is illustrated in Fig.~\ref{fig:diagram}(a). Multichannel mixture signals are transformed between the time domain and the time-frequency domain via STFT/iSTFT. The pipeline consists of three key stages. 
First, the single-channel mask estimation module accepts a single-channel signal and outputs the time-frequency magnitude mask. Without the loss of generality, we select the signal on the first channel to predict the above single-channel mask, which is applied to each channel to obtain estimated speech and noise signals.
Then, we compute ISCMs for speech and noise.
Taking speech as an example, the ISCM is calculated as $\hat{\mathbf{\Psi}}_{\mathbf{xx}}(f, t) = \hat{\mathbf{x}}(f, t) \hat{\mathbf{x}}^H(f, t)$, where $\hat{\mathbf{x}}(f, t)$ is the separated speech signal by single-channel masking.
ISCMs of speech and noise are computed similarly, with different masked outputs.
The core of the second part is attention-based SCM estimation. As shown in Fig.~\ref{fig:diagram}(b), two types of attention-based SCM estimation modules are available: LA and NLA. The final stage computes the spatial filters. The entire system is trained in an end-to-end manner and achieves real-time causal processing during inference.

During the training phase, an oracle magnitude mask \cite{erdogan2015phase} is utilized in the mask estimation module. For inference, a pre-trained causal version of Conv-TasNet model \cite{luo2019conv} applied on the STFT representation is employed. The masking operation is element-wise multiplication. The single-channel masks used are real-valued masks, with values between $0$ and $1$.
\subsubsection{LA and NLA modules}
\label{sssec:LA_NLA}
Before entering the LA and NLA modules, the estimated ISCMs are vectorized row-wise into an one-dimensional vector. Subsequently, the real and imaginary components are represented separately and concatenated to form the real-valued vectorized ISCM. 
The output of the LA and NLA modules needs to be reshaped with the opposite operation as the input process. The estimated vectorized SCM output is first converted into a complex vector and then row-wise stacked into a complex-valued matrix-shaped estimated SCM.

The LA module is designed to estimate attention weights ${w}_{x}(t,\tau)$, which are used to linearly combine ISCMs to estimate the SCMs as
\begin{equation}
\label{eq:L-ACDE}
\hat{\mathbf{\Phi}}_{\mathbf{xx}}(f,t) = \sum_{\tau=1}^{t} {w}_{x}(t,\tau) \hat{\mathbf{\Psi}}_{\mathbf{xx}}(f,\tau).
\end{equation}
The attention weights specify which frames to emphasize when computing the SCM at a given time frame (i.e., $t$) across all past time frames (i.e., $\tau$ = $1, \ldots, t$).
The output and input of the LA module maintain a strict linear relationship, which is also the reason for the naming.
As shown in Fig.~\ref{fig:diagram}(b), the main processing flow of the LA module consists of four parts: a linear layer for reducing the dimensionality of the input vectors, a positional encoding, a stack of core attention blocks, and a dot product computation. 
The linear layer is optional but can effectively reduce the vector length, lessening redundant information and the number of network parameters. The positional encoding is responsible for marking the temporal order of the input vector sequence. Stacking $N$ identical transformer-encoders proposed in \cite{vaswani2017attention} results in the attention blocks ($N$) shown in Fig.~\ref{fig:diagram}(b). The transformer-encoder uses two sub-layers: multi-head attention and a fully connected feed-forward layer. A residual connection and layer normalization are applied after each sub-layer. The scaled dot-product is performed as in \cite{vaswani2017attention}.

The NLA and the LA fundamentally differ in employing the attention mechanism, as the NLA is used to estimate the SCMs directly instead of computing a linear combination of ISCMs.
Similarly to the LA, the NLA also employs the positional encoding and attention blocks ($N$). There are two differences from a network architecture perspective:
1) In NLA, attention scores are no longer calculated after the $N$-th attention block. The output of the $N$-th attention block is used as the output of the NLA or fed to the following linear layer.
2) If dimensionality reduction is performed using a linear layer in NLA, it is mandatory to restore the vector length with the opposite dimension setting after the attention blocks. Regardless of whether a linear layer is used, the input and output of the NLA no longer maintain a linear relationship.

To achieve a real-time causal system, a lower triangular mask is utilized during the attention computation in the attention blocks ($N$) to eliminate information from future frames than the target one \cite{vaswani2017attention}.
When the LA module is adopted, substituting the estimated $\hat{\mathbf{\Phi}}_{\mathbf{xx}}(f,t)$ and $\hat{\mathbf{\Phi}}_{\mathbf{nn}}(f,t)$ from \eqref{eq:L-ACDE} into \eqref{eq:MVDR-filter} yields the LA-MVDR. Similarly, with estimated SCMs from the NLA module and the same substitution, the NLA-MVDR is derived.
\begin{table}[tb]
\centering
\caption{Data and Network Parameter Settings}
\vspace{6pt}
\label{tab:parameter-settings}
\begin{tabular}{p{6.4cm}|c}
\hline
Room Length (m) & [4.0, 8.0] \\
Room Width (m) & [4.0, 8.0] \\
Room Height (m) & [3.0, 4.0] \\
RT60 (s) & [0.3, 0.6] \\
Mic-Array Height (m) & [1.0, 1.5] \\
Min Mic-Array Distance from Wall/Floor (m) & 0.5 \\
Sound Source Height (m) & [1.5, 2.0] \\
Min Sound Source Distance from Wall/Floor (m) & 0.5 \\
Min Mic-Array Distance from Sound Source (m) & 0.2 \\
Number of Movement Trajectories & 50 \\
Sound Source Movement Speed (m/s) & [1.0, 1.5] \\
SNR (dB) & [0.0, 10.0] \\
\hline
Batch size & 8 \\
Learning rate & $1 \times 10^{-4}$ \\
Number of attention blocks ($N$) & 2 \\
Number of attention heads & 4 \\
Dimension of attention layers & 256 \\
Dimension of feed-forward layers & 2048 \\
\hline
\end{tabular}
\vspace{-2ex}
\end{table}
\subsection{Attention-based inverse-covariance estimation}
\label{ssec:filter}
We can relax the limitations in \eqref{eq:MVDR-filter}, removing the need for explicit matrix inversion or trace calculations. Here, we use the NLA to directly estimate the inverses of the covariance matrices, resulting in the inverse-covariance MVDR (IC-MVDR), which is calculated as
\begin{equation}
\label{eq:NLA-IF}
\mathbf{h}_{\text{IC-MVDR}}(f,t) = \mathbf{A}_{\mathbf{xx}, \text{NLA}}(f,t) \mathbf{A}_{\mathbf{nn}, \text{NLA}}(f,t) \mathbf{u}_{\text{ref}},
\end{equation}
where $\mathbf{A}_{\mathbf{xx}, \text{NLA}}(f,t)$ and $\mathbf{A}_{\mathbf{nn}, \text{NLA}}(f,t)$ are estimates at the time-frequency bin $(f, t)$ obtained by the NLA module. 
In this case, the NLA is used to estimate the scaled inverse of the SCMs instead of the SCMs.
\subsection{Direct spatial filter estimation}
\label{ssec:FL-MCLF}
The fully learnable spatial filter (FL-SF) is implemented using the framework shown in Fig.~\ref{fig:diagram}(c). This approach utilizes the NLA to directly estimate a spatial filter. The complex-valued spectrograms of the mixture signal obtained after STFT are vectorized row-wise into a one-dimensional vector on each frame, and then the real and imaginary parts are extracted and concatenated into real-valued vectorized spectrograms, and fed to the NLA.
The output of the NLA, once converted into complex vectors and stacked into a complex-valued matrix, gives directly the desired time-varying spatial filter $\mathbf{h}_{\text{FL-SF}}$.
The enhanced signal is then obtained by iSTFT after linear filtering in \eqref{eq:multi-channel-filtering}.
\subsection{Averaging-based SCM estimation}
\label{ssec:HB-MCLF}
As a baseline in the experiments, we use conventional spatial filtering methods where the SCMs are estimated by averaging ISCMs. The three alternative strategies are described below using the speech signal as an example.
1) Cumulative averaging (CUM-AVG)
\begin{equation}
\label{eq:mean-SCM}
\hat{\mathbf{\Phi}}_{\mathbf{xx}}(f,t) = \frac{1}{t} \sum_{\tau=1}^{t} \hat{\mathbf{\Psi}}_{\mathbf{xx}}(f,\tau)
\end{equation}
weights equally all frames of the whole utterance.
2) Recursive averaging (REC-AVG)
\begin{equation}
\label{eq:recursive-SCM}
\hat{\mathbf{\Phi}}_{\mathbf{xx}}(f,t) = \alpha \hat{\mathbf{\Phi}}_{\mathbf{xx}}(f,t-1) + \hat{\mathbf{\Psi}}_{\mathbf{xx}}(f,t)
\end{equation}
uses first-order recursive filtering with forgetting factor $\alpha$.
3) For block-wise averaging (BLOCK-AVG), $W$ latest frames are averaged as
\begin{equation}
\label{eq:moving-ave-SCM}
\hat{\mathbf{\Phi}}_{\mathbf{xx}}(f,t) = \frac{1}{W} \sum_{\tau=t-W+1}^{t} \hat{\mathbf{\Psi}}_{\mathbf{xx}}(f,\tau).
\end{equation}
In the experiments, a forgetting factor of $0.95$ was applied in \eqref{eq:recursive-SCM}, while a window length of $25$ frames (i.e., $400$ ms) was used in \eqref{eq:moving-ave-SCM}. The parameters were set from multiple tests on the validation set. 
\subsection{Network complexity optimization}
\label{ssec:complexity-optimization}
The parameter size of the DNNs is a significant factor in real-world applications. We use three methods to substantially reduce the parameter size in the DNN-integrated spatial filtering system without compromising performance significantly.
1) The ISCMs retain only the diagonal and the lower triangular parts. In SCM matrix reconstruction, the process is reversed.
Since the ISCM is a Hermitian matrix, the upper triangular part is reconstructed based on the lower triangular part.
2) A single LA or NLA can be used for speech and noise SCM estimation with different masked inputs.
3) The linear layers in both LA and NLA modules can reduce the dimensionality of the input.

  
  

\section{Evaluation}
\label{sec:experiments}
\subsection{Dataset}
\label{ssec:data}
To assess our system, a $5$-channel dataset was synthesized. Speech signals were from \textit{WSJ0} \cite{paul1992design} and real-world noises were sourced from \textit{CHiME-3} \cite{barker2015chime}. The synthesis parameters are presented in Table~\ref{tab:parameter-settings}. The process comprised three primary steps:

Step 1: Trajectories of the dynamic sound source and microphone array positions were generated randomly according to the specified parameters listed in Table~\ref{tab:parameter-settings}, e.g., the source movement speed was between $1.0$ m/s and $1.5$ m/s.
    
Step 2: Utilizing the \textit{gpuRIR} toolbox \cite{diaz2021gpurir}, multichannel reverberant speech signals were simulated based on the randomly generated spatial parameters and speech segments from \textit{WSJ0}.
    
Step 3: Noise segments, sourced from \textit{CHiME-3}, were scaled by the predefined SNR values in Table~\ref{tab:parameter-settings} and then mixed with the speech signal to yield the final mixture signals.

    
    

The \textit{CHiME-3} employs a $6$-element planar microphone array, where the orientation of the second microphone contrasts with the remaining five. During synthesis, the channel of this particular microphone was excluded. The spatial position of the microphone array was randomly determined, focusing only on its spatial displacement without rotations. Synthesized signals span between $1$ and $15$ seconds, sampled at $16$ kHz. Segments are sampled randomly from \textit{WSJ0} and \textit{CHiME-3} with no overlaps among the training, validation, and test sets. The resulting dataset comprises $20000$ training samples,  $2000$ validation samples, and $2000$ test samples.

A 5-channel static dataset was synthesized using identical parameters. The only difference is that we only considered the first positional point from the randomly generated spatial trajectories of moving sound sources in step 2.

\begin{table}[tb]
\centering
\caption{Experimental results in static situations}
\vspace{2ex}
\label{tab:Experimental-results-static} 
\renewcommand{\arraystretch}{1.1}
\begin{tabular}{c|ccc}
\hline
\multirow{2}{*}{Methods} 
& \multicolumn{3}{c}{Static sources} \\
\cline{2-4}
& SDR & PESQ & STOI \\
\hline
LA-MVDR & 12.9 & 2.32 & 0.94 \\
NLA-MVDR & 12.9 & \textbf{2.36} & \textbf{0.95} \\
IC-MVDR & \textbf{13.2} & 2.27 & 0.94 \\
\hline
FL-SF & 12.2 & 2.16 & 0.91  \\
\hline
CUM-AVG-MVDR & 12.2 & 2.11 & 0.91 \\
REC-AVG-MVDR & 10.9 & 2.07 & 0.92 \\
BLOCK-AVG-MVDR & 11.2 & 2.09 & 0.92 \\
\hline
\end{tabular}
\vspace{-2ex}
\end{table}

\begin{table}[tb]
\centering
\caption{Experimental results in dynamic situations}
\vspace{2ex}
\label{tab:Experimental-results-dynamic} 
\renewcommand{\arraystretch}{1.1}
\begin{tabular}{c|ccc}
\hline
\multirow{2}{*}{Methods} 
& \multicolumn{3}{c}{Dynamic sources} \\
\cline{2-4}
& SDR & PESQ & STOI \\
\hline
LA-MVDR & 12.8 & \textbf{2.31} & \textbf{0.94} \\
NLA-MVDR & 12.4 & 2.19 & 0.92 \\
IC-MVDR & \textbf{12.9} & 2.24 & 0.93 \\
\hline
FL-SF & 11.8 & 2.10 & 0.90  \\
\hline
CUM-AVG-MVDR & 9.1 & 1.95 & 0.89 \\
REC-AVG-MVDR & 10.1 & 2.06 & 0.91 \\
BLOCK-AVG-MVDR & 9.7 & 2.03 & 0.90 \\
\hline
\end{tabular}
\vspace{-1ex}
\end{table}
\subsection{Experiment configurations}
\label{ssec:Experiment-configurations}
The Hanning window and STFT length were both configured at 1024 samples, with a hop length of 256 samples. In the training of Conv-TasNet, hyperparameters were defined as follows: \( B=256 \), \( H=512 \), \( X=8 \), \( R=4 \), and \( Sc=256 \). A sigmoid function was employed for mask activation. The Adam optimizer was utilized with a batch size of $16$ and a learning rate of \(1 \times 10^{-4}\). Throughout the training, the learning rate scheduler and early-stopping methodologies were implemented. The parameters were chosen with reference to \cite{luo2019conv}.
The parameters for LA and NLA are specified in Table~\ref{tab:parameter-settings} by referencing established practices in \cite{vaswani2017attention,ochiai2023mask}. Adam optimized was used. The learning rate scheduler and early-stopping were adopted.
\subsection{Training objective and evaluation metrics}
\label{ssec:loss-metrics}
We use a negative utterance-level signal-to-noise ratio (SNR) as the loss function for end-to-end training, defined as
\vspace{1ex}
\begin{equation}
\label{eq:loss-function}
\mathcal{L}(\mathbf{s}, \hat{\mathbf{s}}) = -10 \cdot \log_{10}\left(\frac{\|\mathbf{s}\|^{2}}{\|\mathbf{s}-\hat{\mathbf{s}}\|^{2}}\right).
\end{equation}
Here $\mathbf{s}$ is the reverberant clean speech signal on the reference microphone, and $\hat{\mathbf{s}}$ is the enhanced signal. Three metrics are utilized for evaluation: SDR, PESQ, and STOI. 
\section{Results and analysis}
\label{sec:results}
Tables~\ref{tab:Experimental-results-static} and \ref{tab:Experimental-results-dynamic} display the results for static and dynamic sound sources, respectively, which are averaged from the test set evaluations.
In Table~\ref{tab:Experimental-results-static}, the IC-MVDR achieves the highest SDR gain, while the NLA-MVDR attains the top PESQ and STOI.
The CUM-AVG-MVDR method outperforms both the REC-AVG-MVDR and BLOCK-AVG-MVDR on the static sound source dataset. The FL-SF performance falls between the two methods above.

In Table~\ref{tab:Experimental-results-dynamic} for dynamic situations, the IC-MVDR still achieves the best SDR performance, whereas the PESQ and STOI performance of LA-MVDR is the best.
Additionally, the REC-AVG-MVDR method is good in the SDR, PESQ, and STOI metrics, while the performance of the CUM-AVG-MVDR declines in comparison to the static cases.
Despite a performance decrease of the FL-SF with moving sound sources, it still outperforms conventional methods

Comparing the static and dynamic results, it is clear that the conventional spatial filters are sensitive to sound source movement, as three weighting strategies exhibit noticeable performance degradation in dynamic environments. However, the attention-driven methods display robust performance in both static and dynamic tests, where the SDR performance degradation is less than $0.5$~dB.

The experimental results also provide some enlightening conclusions:
1) Employing DNNs for estimating time-varying statistics in the conventional multichannel spatial filtering pipeline, such as MVDR, results in filters with speech quality and intelligibility advantages compared to the fully learnable spatial filters. One possible reason is that the derivation of conventional spatial filters considers prior information from the signal model and reduces speech distortion.
2) Using DNNs to directly estimate the inverses of the covariance matrices improved the SDR performance of the resulting filters.
3) The fully learnable spatial filter results showed that the errors from a spatial filtering system can be optimized in a single step, thereby avoiding the multiple errors introduced within different stages in the widely used multi-stage frameworks, such as mask-based neural beamforming.  

\section{Conclusions}
\label{sec:conclusions}
This paper investigated attention-based SCM estimation methods and a fully learnable spatial filter for multichannel speech enhancement. The attention-driven approaches showed strong robustness across dynamic and static environments, outperforming conventional spatial filtering techniques. 
We proposed a method that implements MVDR by estimating the inverses of covariances matrices, showing a clear SDR performance improvement over two kinds of datasets. Additionally, the approach of weighting ISCMs using attention weights to estimate SCMs outperformed the reference methods in terms of PESQ and STOI.
Incorporating the attention mechanism into a conventional spatial filtering pipeline significantly improves the overall performance compared to conventional averaging-based methods.

\section{Acknowledgment}
The research was funded by and conducted in collaboration with Nokia Technologies.

\newpage



\end{sloppy}

\end{document}